\documentclass[aps, prl, twocolumn, superscriptaddress, amsmath, amssymb, floatfix,nolongbibliography, nofootinbib, notitlepage]{revtex4-2}
\usepackage{amsbsy}
\usepackage{amsfonts}
\usepackage{amsmath}
\usepackage{amstext}
\usepackage{amsthm}
\usepackage{amssymb}

\usepackage{graphicx}

\usepackage{epsfig}
\usepackage{epstopdf}
\usepackage{bm}
\usepackage{color}
\definecolor{prlblue}{rgb}{0.18,0.19,0.57}
\usepackage{multirow}
\usepackage[colorlinks]{hyperref}
\usepackage{cleveref}
\hypersetup{citecolor=prlblue, linkcolor=prlblue, filecolor=prlblue, urlcolor=prlblue}
\newcommand{\diag}[2]{\vcenter{\hbox{\includegraphics[height=#2]{#1}}}}
\newcommand{\figref}[1]{Fig.\,\ref{#1}}

\newcommand{\eqnref}[1]{Eq.\,\eqref{#1}}

\renewcommand{\Im}{\mathop{\mathrm{Im}}}

\begin{document}
\graphicspath{{figures/}}

%\title{Two-component Spinon Description in Doped Mott Insulator}
\title{Hourglass-Like Spin Excitation in a Doped Mott Insulator}

\author{Jia-Xin Zhang}
\thanks{These authors contributed equally to this work}
\email{zjx19@mails.tsinghua.edu.cn}
\affiliation{Institute for Advanced Study, Tsinghua University, Beijing 100084, China}

\author{Chuan Chen}
\thanks{These authors contributed equally to this work}
\affiliation{Institute for Advanced Study, Tsinghua University, Beijing 100084, China}

\author{Jian-Hao Zhang}
\affiliation{Department of Physics, The Pennsylvania State University, University Park, Pennsylvania 16802, USA}
\author{Zheng-Yu Weng}
%\email{weng@mail.tsinghua.edu.cn}
\affiliation{Institute for Advanced Study, Tsinghua University, Beijing 100084, China}

\date{\today}
%%%%%%%%%%%%%%%%%%%%%%%%%%%%%%%%%%%%%%%%%%%%%%%%%%%%%%%%%%%%%%%%%%%%%%%%%%%%%%%%%%%%%%%%%%%%%%

\begin{abstract}
We examine the dynamical magnetic response in a two-component resonating-valence-bond (RVB) description of the doped Mott insulator. The half-filled antiferromagnetic phase described by the Schwinger-boson mean-field theory will evolve into a bosonic-RVB state in the superconducting phase upon doping, where the doped holes introduce another fermionic itinerant spinon which forms a BCS-like RVB order. The spin excitations are thus composed of a resonance-like mode from the former and a weak dispersive mode from the itinerant component at the mean-field level. These two-component spinons are shown to give rise to an hourglass-like spin excitation at the RPA level via an antiferromagnetic coupling between the two modes, which provides an unconventional explanation of the experimental observations in the cuprate. In particular, we also discuss an instability towards an incommensurate magnetic order in this theoretical framework.

\end{abstract}

\maketitle

%\tableofcontents

%%%%%%%%%%%%%%%%%%%%%%%%%%%%%%%%%%%%%%%%%%%%%%%%%%%%%%%%%%%%%%%%%%%%%%%%%%%%%%%%%%%%%%%%%%%%%%
%%%%%%Introduction
\begin{figure}[htb]
	\centering
	\includegraphics[width=\linewidth]{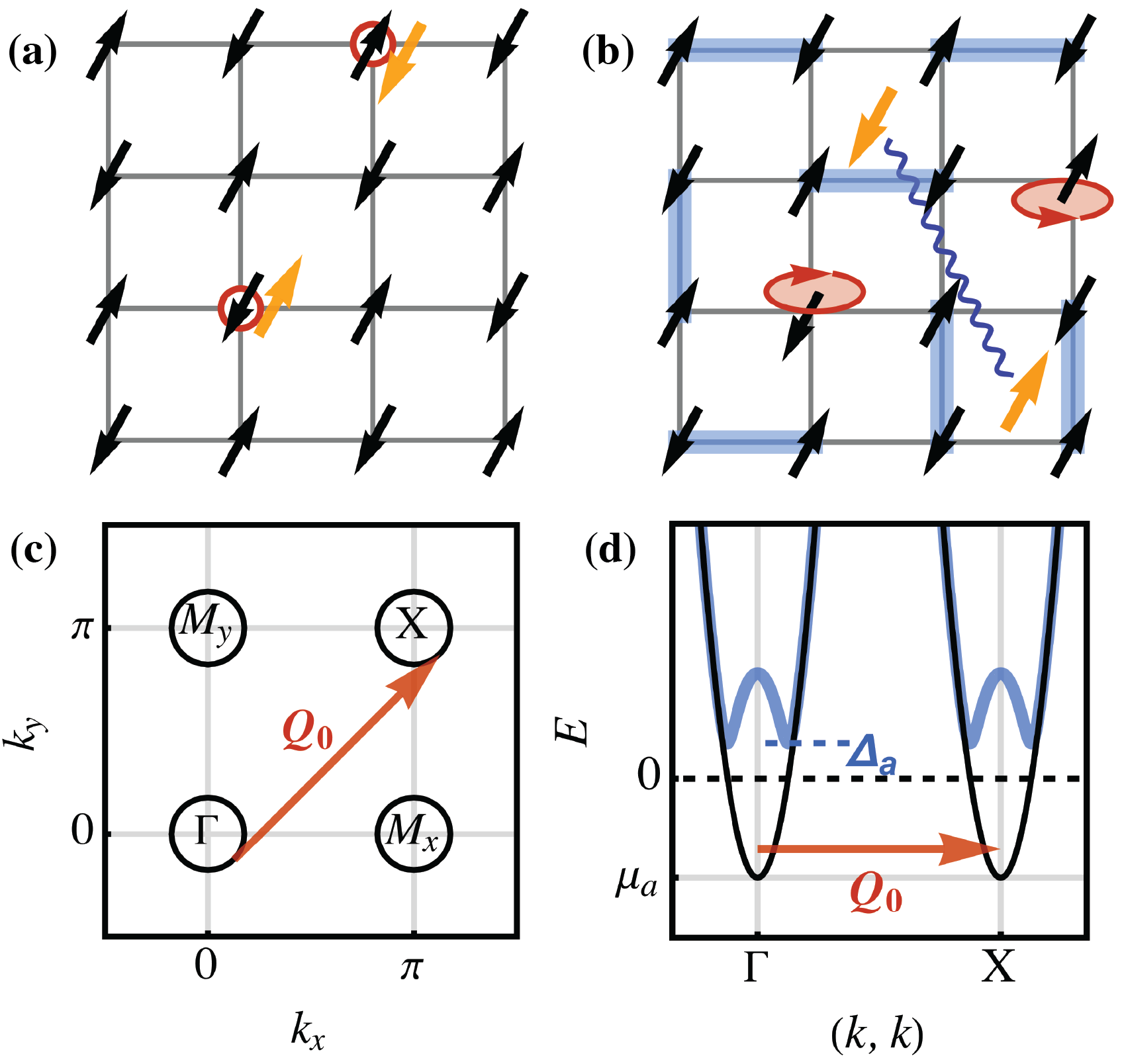}
	\caption{Schematic illustration of the two-component spinons in a doped Mott insulator. (a) A bare hole is composed of a bosonic holon (red circle) and a fermionic $a$-spinon (orange arrow) in a spin background filled with the single-occupied bosonic $b$-spinons (black arrow) such that the total spin at the hole site is zero; (b) Two-component RVB state in which holons are condensed and $b$-spinons form singlet RVB pairings (blue lines), with each unpaired $b$-spinon carrying a $\pi$-vortex (red circle with arrow) of the charge supercurrent. Concurrently, the $a$-spinons are in an $s$-wave pairing (wavy lines); (c) Four Fermi pockets for the $a$-spinons emerge if the pairing order parameter $\Delta_a$ vanish. The red arrow denotes the AFM wavevector $\boldsymbol{Q}_0=(\pi,\pi)$; (d) Energy dispersion of the $a$-spinon near $\Gamma$ and $X$ pockets displayed by black curves for $\Delta_a=0$ and blue curves for $\Delta_a\neq 0$.}
	\label{fig_GS}
\end{figure}
{\it Introduction.---}The spin dynamics is essential for understanding the mechanism of the cuprate superconductor, which reduces to the only relevant low-lying mode in the undoped limit \cite{Wen.Lee.2006z4}. At finite doping, the dynamic spin susceptibility measured by the inelastic neutron scattering (INS) reveals that the gapless spin-wave \cite{Fisk.Coldea.2001, Perring.Headings.2010} at the antiferromagnetic (AFM) wave vector $\boldsymbol{Q}_0=(\pi,\pi)$ becomes gapped with the destruction of the AFM long-range order. The spin excitation further displays a resonance-like mode \cite{PhysRevLett.75.316, Keimer.He.20006wc, Bourges.Fauqu.2007, Keimer.Capogna.2007, Keimer.Fong.1999, Keimer.He.2002, Dai1999} with a characteristic energy $E_g$. Slightly deviating from $\boldsymbol{Q}_0$, the resonance mode splits and extends to both higher and lower energies to result in the well-known hourglass-shaped spectrum \cite{Greven.Chan.2016, Greven.Chan.2016noa, Keimer.Pailhès.2004, Keimer.Hinkov.2007, Tranquada.Xu.2009, Perring.Vignolle.2007, Dogan.Hayden.2004, Yamada.Tranquada.2004, Fujita.Sato.2020}. 

Phenomenologically, two distinct starting points have been commonly employed to describe the experimentally observed dynamical spin susceptibility. One is based on the itinerant magnetism approach \cite{Pfeuty.Onufrieva.2002, Norman.Eremin.2004, Schmalian.Abanov.2002}, where the spin resonance formation below $T_c$ originates from the enhanced feedback effect of the $d$-wave superconductivity for quasiparticles with a large Fermi surface. Alternatively, the local moment approach \cite{Schreiber.Sherman.2003, Eremin.Eremin.2012, Rice.James.2012, Sushkov.Milstein.2008} starts with the undoped two-dimensional (2D) AFM state by examining a mixture of local spins described by the superexchange interaction $J$ and itinerant carriers with tight-binding energy dispersion. 

%%%the 2D copper-oxide layer in the parent compound of the cuprate superconductor
Microscopically, the parent compound of the cuprate acts as a Mott insulator, in which all the electrons form local magnetic moments as described by the minimal AFM Heisenberg model at half-filling. How such an AFM state can be doped into a short-range AF state at finite doping has been a central issue in the study of the doped Mott insulator, which is described by an effective one-band model, e.g., the $t$-$J$ model \cite{Anderson.Anderson.1987, Rice.Zhang.1988}. The fermionic RVB state was originally proposed by Anderson \cite{Anderson.Baskaran.1987, Anderson.Anderson.1987} is one of the conjectures for such a phase, which results in a d-wave Superconducting (SC) instability at low temperatures \cite{Shiba.Zhang.1988, Liu.Kotliar.1988}. Nevertheless, this fermionic RVB state seems incompatible with the Schwinger-boson or bosonic RVB description\cite{ Lee.Chakraborty.1990, Auerbach.Arovas.1988, Wen.Lee.2006z4, Read.Sachdev.1991} of the AFM state at half-filling, and how to bridge the two phases still remains unclear \cite{Nagaosa.Lee.1992, Wen.Lee.2006z4}.
Recently, a two-component RVB description has been proposed\cite{Weng.Weng.2011,Weng.Ma.2014, Zhang.Weng_2022}, which theorizes doping an AFM state into a short-range AF state with an intrinsic low-temperature SC instability. Here the AFM phase is well characterized by the Schwinger-boson mean-field state at half-filling, which is then turned into a bosonic RVB state by doping due to the phase-string effect\cite{Weng.Sheng.1996,Weng.Ma.2014} generally associated with a doped Mott insulator. The latter will lead to a nontrivial spin-current backflow created by doped holes moving in a spin singlet background\cite{Weng.Zheng.2018, Weng.Zhao.2022}. The resulting spin current, in combination with the doped holes, gives rise to distinct spinons which are fermionic and itinerant in nature\cite{Weng.Ma.2014,Zhang.Weng_2022}.

In this paper, we study an unconventional spin excitation in the doped Mott insulator at finite doping as the consequence of such a two-component RVB description.  At the RPA level, such a new spin excitation is hourglass-like, which is composed of the bosonic spinons evolved from the Schwinger bosons at half-filling and the itinerant fermionic spinons emerging upon doping. The result is consistent with the INS observations\cite{PhysRevLett.75.316, Keimer.He.20006wc, Bourges.Fauqu.2007, Keimer.Capogna.2007, Keimer.Fong.1999, Keimer.He.2002, Dai1999} in the cuprate.  Further physical implications are also discussed.

%%%%%%%%%% two-component RVB state%%%%%%%%%%%%%%
{\it Emergent two-component RVB description at finite doping.---} Starting from the half-filling by doping, a two-component RVB description of the short-range AF state has been recently proposed\cite{Weng.Weng.2011,Weng.Ma.2014} based on the $t$-$J$ model, whose ground state is given by
\begin{equation}\label{eq:gs}
    |\Psi_G\rangle={\hat {\cal P}}\left[e^{i\hat{\Theta}}|\Phi_h\rangle  \otimes|\Phi_a\rangle \otimes|\Phi_b\rangle \right] ~.
\end{equation}
Here $|\Phi_b\rangle$ originated from the Schwinger-boson mean-field state at half-filling and is known as the bosonic RVB state[shown by blue thick lines in \figref{fig_GS}(b)], $|\Phi_a\rangle $ is a BCS-like state[shown by blue wave lines in \figref{fig_GS}(b)] formed by the \emph{fermionic} spinons which are introduced by the doped holes, and $|\Phi_h\rangle$ describes a Bose-condensed state of the bosonic holons which are also introduced by the doped holes as carrying electric charges.

The unitary operator $e^{i\hat{\Theta}}$ in Eq. (\ref{eq:gs}) is a duality transformation to implement the so-called phase-string effect\cite{Weng.Sheng.1996, Weng.Ma.2014}, which is very singular as created by the doped holes. The projection operator ${\hat {\cal P}}$ further enforces the constraint between the three fractionalized sub-systems in Eq. (\ref{eq:gs}) by
\begin{equation}\label{Sab}
	n_i^h S_b^z(\boldsymbol{r}_i)=-S_a^z(\boldsymbol{r}_i),
\end{equation}
in which $n_i^h$ is the holon number at site $i$, and  $S_a^z$ and $S_b^z$ denote the $z$-component spins of the $a$-spinon and $b$-spinon, respectively. Physically, Eq.(\ref{Sab}) means the half-filled $b$-spinons at the hole sites must be compensated by the $a$-spinons, whose number is equal to the hole number[depicted in \figref{fig_GS}(a)]. Previously, the individual behaviors for $|\Phi_b\rangle$, and $|\Phi_a\rangle$ have been studied\cite{Weng.Ma.2014, Zhang.Weng_2022, Weng.Mei.20107w, Weng.Chen.2005}, whose results will be first given in the following. Then the effect of ${\hat {\cal P}}$ in Eq. (\ref{Sab}) will be further incorporated at the RPA level.

\begin{figure}[t]
\centering
\includegraphics[width=0.99\linewidth]{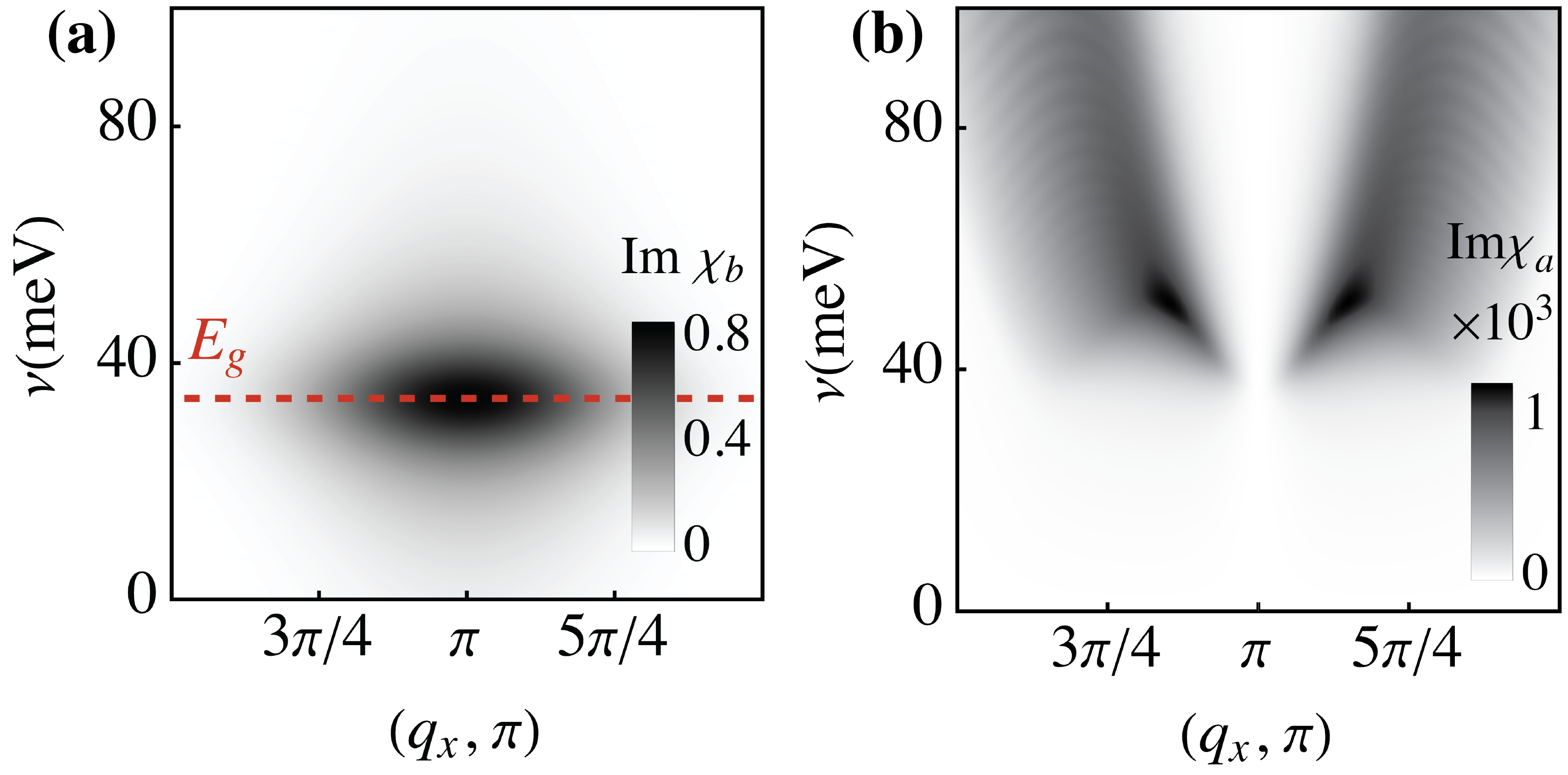}
\caption{(a) Imaginary part of bare dynamic spin susceptibility $\operatorname{Im} \chi_b \left(q\right)$ for $b$-spinons, derived from \eqnref{chib} near the AFM wave vector $\boldsymbol{Q}_0$ at $\delta=0.1$, with the red dashed line indicating the resonance energy $E_g$. (b) Corresponding susceptibility $\operatorname{Im} \chi_a \left(q\right)$ for $a$-spinons, obtained from \eqnref{chia}. Parameter values are provided in the main text.}
\label{fig_chi}
\end{figure}

{\it Local moments.---} At half-filling, the ground state of the Heisenberg Hamiltonian is well described by the Schwinger-boson mean-field state\cite{Lee.Chakraborty.1990, Auerbach.Arovas.1988, Wen.Lee.2006z4, Read.Sachdev.1991}, which will evolve into the short-range AF state $|\Phi_b\rangle$ at finite doping as outlined above[cf. blue thick line in \figref{fig_GS}(b)]. In contrast to conventional Schwinger bosons with continuous spectra \cite{Auerbach.Arovas.1988}, the $b$-spinons in this study exhibit dispersionless, ``Landau-level-like'' discrete energy levels with a gap $E_s$ \cite{SM, Zhang.Weng_2022, Weng.Chen.2005}. Consequently, the corresponding low-lying dynamical spin susceptibility originating from the lowest Landau level is given by as \cite{SM,Weng.Chen.2005, Weng.Mei.20107w, Zhang.Weng_2022}

\begin{eqnarray}\label{chib}
\chi_b\left(i\nu_n,\boldsymbol{Q}\right)=\diag{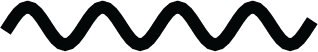}{4pt}&=&a_c^2 D e^{-\frac{a_c^2}{2}\left(\boldsymbol{Q}-\boldsymbol{Q}_0\right)^2}  \\
    &\;& \times \left(\frac{1}{i\Omega_n-E_g}-\frac{1}{i\Omega_n+E_g}\right),\notag
\end{eqnarray}
where $E_g=2E_s$ represents the resonance energy, the ``cyclotron length'' $a_c=a / \sqrt{\pi \delta}$ determines the effective spin-spin correlation length[$a$ for lattice constant, $\delta$ for doped hole density], and the weight $D$ is not sensitive to doping \cite{SM}. As depicted in \figref{fig_chi}(a), the spin-wave excitation, derived from the imaginary component of \eqnref{chib}, becomes a gapped resonance-like mode near $\boldsymbol{Q}_0=(\pi,\pi)$.

%%%%%%itinerant spinons
{\it Itinerant spinons.---}
The doped holes are created by removing spins from the half-filling spin-singlet background characterized by $|\Phi_b\rangle$. The doping introduces new spinons centered at the hole sites known as the $a$-spinons [the yellow arrows in \figref{fig_GS}(a)], which form the itinerant RVB state $|\Phi_a\rangle$ in Eq. (\ref{eq:gs}) [cf. blue wave line in \figref{fig_GS}(b)]. 

The $a$-spinons as fermions form the multi-pocket Fermi surfaces illustrated in \figref{fig_GS}(c), which are determined by:
\begin{eqnarray}\label{Ha}
H_a&=&\sum_{\boldsymbol{K}, \boldsymbol{k}} \epsilon_{\boldsymbol{K}}(\boldsymbol{k}) a_{\boldsymbol{K}+\boldsymbol{k}, \sigma}^{\dagger} a_{\boldsymbol{K}+\boldsymbol{k}, \sigma} \notag \\
&\;&+\sum_{\boldsymbol{K}, \boldsymbol{k}} \Delta_a a_{\boldsymbol{K}+\boldsymbol{k}, \uparrow}^{\dagger} a_{\boldsymbol{K}-\boldsymbol{k}, \downarrow}^{\dagger}+\text { h.c. }.
\end{eqnarray}
Here $a_{\boldsymbol{K}+k, \sigma}^\dagger$ denotes the creation operator for an itinerant $a$-spinons from pockets $\boldsymbol{K}=\Gamma, X, M$ with relative momentum $\boldsymbol{k}$[depicted in \figref{fig_GS}(c)], whose band energy reads $\epsilon_K(\boldsymbol{k})=\boldsymbol{k}^2/2 m_a-\mu_a$. The $\Delta_a$ term characterizes the uniform $s$-wave pairing within all pockets. We also assume identical parabolic band structures for all pockets as shown in \figref{fig_GS}(d), implying a consistent effective mass $m_a$ and chemical potential $\mu_a$. This model aligns with hopping fermions in the $\pi$ flux states, displaying well-nested, distinct pockets \cite{SM,Weng.Ma.2014, Weng.Zhang.2020, Zhang.Weng_2022}. Importantly,  the Luttinger sum rule for itinerant $a$-spinons, which arise from doped holes, is associated with the doping density $\delta$, represented as $\sum_{\boldsymbol{k},\sigma}n_{\boldsymbol{k},\sigma}^a/N=\delta$ [where $n_{\boldsymbol{k},\sigma}^a$ denotes the $a$-spinon number operator and $N$ denotes the total number of sites], rather than half-filling as in conventional spin liquids \cite{Ng.Zhou.2017}. This relationship determines the chemical potential $\mu_a$.

The dynamical spin susceptibility of itinerant $a$-spinons is defined as $\chi_a\left(r_i-r_j\right)=\left\langle S_a^z\left(r_i\right) S_{a}^z\left(r_j\right)\right\rangle$, with $r_i=(\tau_i, \boldsymbol{r}_i)$ representing the time-space vector. The $\chi_a$ can be formulated in the frequency-momentum space as:

\begin{eqnarray}\label{chia}
    &\;&\chi_a(i v_{n},\boldsymbol{q})=\diag{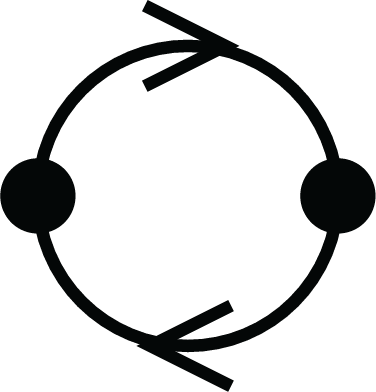}{24pt}=-\frac{1}{2 N} \sum_{\boldsymbol{k}}\left(1-\frac{\Delta_a^2+\epsilon_{\boldsymbol{k}+\boldsymbol{q}} \epsilon_{\boldsymbol{k}}}{E_{\boldsymbol{k}+\boldsymbol{q}} E_{\boldsymbol{k}}}\right) \notag  \\
    &\;&\;\;\;\;\;\;\times \left(\frac{1}{i v_n-E_{\boldsymbol{k}+\boldsymbol{q}}-E_{\boldsymbol{k}}}-\frac{1}{i v_n+E_{\boldsymbol{k}+\boldsymbol{q}}+E_{\boldsymbol{k}}}\right),
\end{eqnarray}
where the term in the first parenthesis represents the coherence factor due to BCS-type pairing and the solid line $\diag{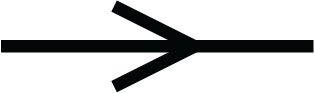}{6pt}$ formally denotes the $a$-spinon propagator. The $\boldsymbol{q}$ in \eqnref{chia} denotes the momentum deviation from all the nesting vectors, such as $(0,0)$, $(\pi,\pi)$, $(0,\pi)$, or $(\pi, 0)$, and it can be easily verified that they are identical. 

The dynamic spin susceptibility is given by $\operatorname{Im} \chi(\nu + i0^+, \boldsymbol{q})$ after the analytic 
continuation $i\nu_n \rightarrow \nu + i0^+$, as depicted in \figref{fig_chi}(b). The spin spectrum around the AFM wave vector $\boldsymbol{Q}_0$, contributed by the scattering between $\Gamma$($M_x$) and $X$($M_y$) pockets, exhibits a continuum above the gap $2\Delta_a$. A significant feature is the complete disappearance of the weight at exact $\boldsymbol{Q}_0=(\pi, \pi)$ due to the coherence factor effect \cite{Aksay.Fong.1995zl, Hu.Seo.2009, Eremin.Korshunov.2008, Scalapino.Maier.2008} of the uniform $s$-wave pairing, i.e., $1-(\Delta_a^2+\epsilon_{\boldsymbol{k}+\boldsymbol{q}} \epsilon_{\boldsymbol{k}})/E_{\boldsymbol{k}+\boldsymbol{q}} E_{\boldsymbol{k}} \xrightarrow[]{\boldsymbol{q}\rightarrow 0} 0$, which is crucial in yielding an ``hourglass'' dispersion in the subsequent results.

{\it Hybrid model.---} So far at the mean-field level, two-component $a$ and $b$ spinons are separated. At the next step, the local spin constraint \eqnref{Sab} will be incorporated at the RPA level via the following local coupling, which is given by:  
\begin{equation}\label{Hint}
    H_{\text{int}}= g \sum_i S_a^z(\boldsymbol{r}_i)  S_b^z(\boldsymbol{r}_i),
\end{equation}
where $g>0$ represents the strength of this effective interaction. At the RPA level, the dynamical spin susceptibility based on \eqnref{Hint} can be diagrammatically expressed as:
\begin{eqnarray}\label{chiRPA}
    \chi^{\mathrm{RPA}}(q)&=&\diag{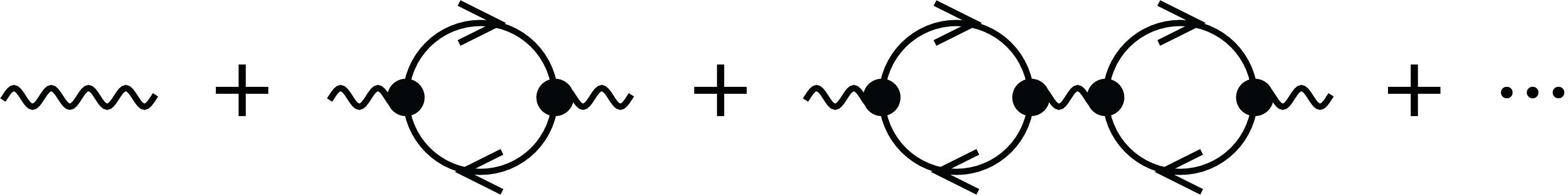}{22pt} \notag \\
    &=&\frac{\chi_b(q)}{1-g^2 \chi_a(q) \chi_b(q)}.
\end{eqnarray}

\begin{figure}[t]
\centering
\includegraphics[width=0.99\linewidth]{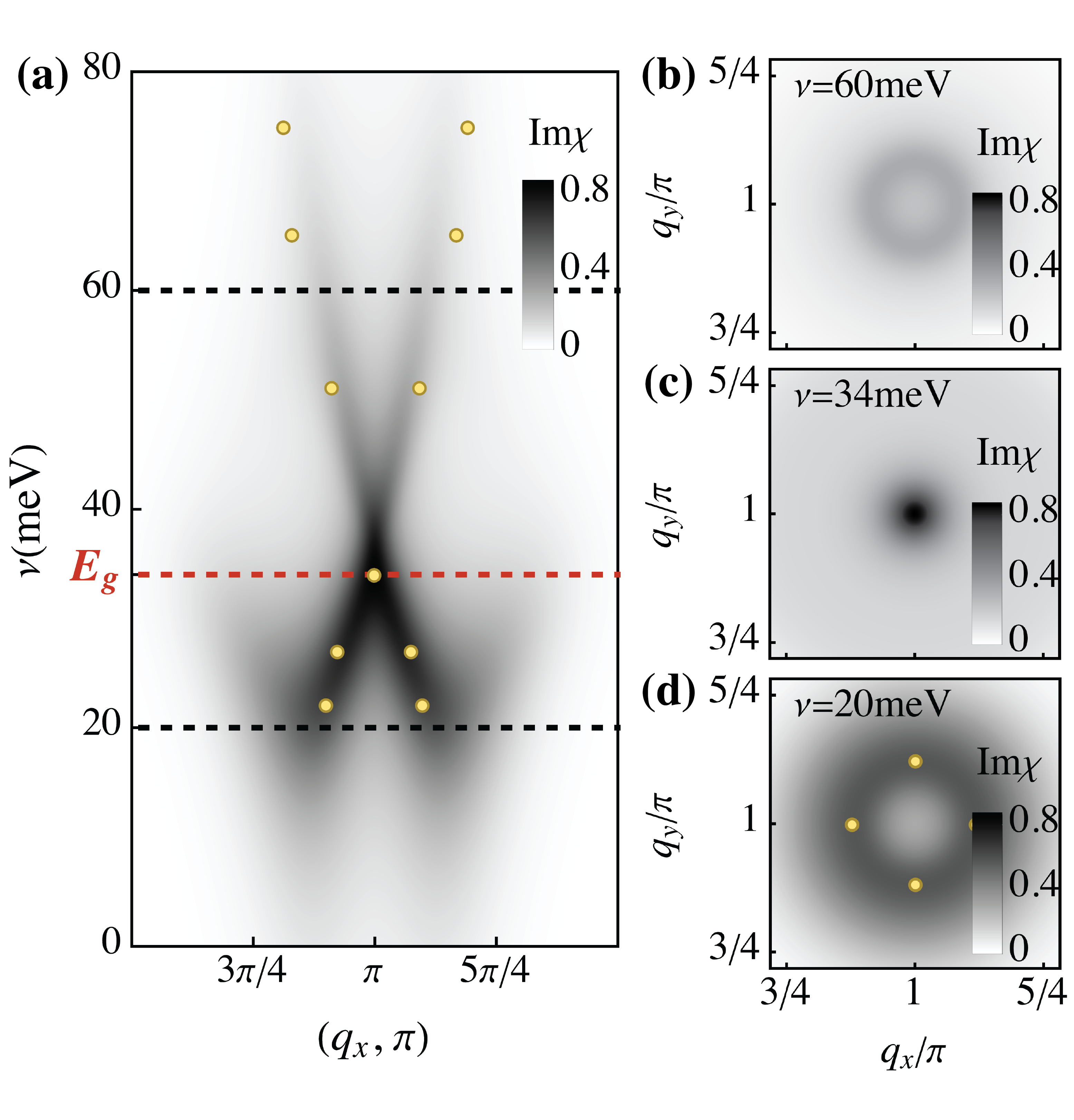}
\caption{(a) Imaginary part of dynamic spin susceptibility at RPA level, $\Im\chi^\text{RPA}(q)$, determined by \eqnref{chiRPA} around AFM wave vector $\boldsymbol{Q}_0$ at $\delta=0.1$ and $g=60 \text{meV}$. (b)-(d) Calculated slices of $\Im \chi^\text{RPA}(q)$ at frequencies indicated by dashed lines in (a). Yellow points in (a) and (d) represent INS results observed in Ref. \onlinecite{Dogan.Hayden.2004}.}
\label{fig_chiRPA}
\end{figure}

The low-energy spin spectrum, $\operatorname{Im} \chi^{\mathrm{RPA}}(q)$, around the AFM wave vector $\boldsymbol{Q}_0$ is depicted in \figref{fig_chiRPA}(a) at $\delta=0.1$, resembling the well-known ``hourglass'' spectrum observed in INS\cite{Greven.Chan.2016, Greven.Chan.2016noa, Keimer.Pailhès.2004, Keimer.Hinkov.2007, Tranquada.Xu.2009, Perring.Vignolle.2007, Dogan.Hayden.2004, Yamada.Tranquada.2004, Fujita.Sato.2020}[with experimental results\cite{Dogan.Hayden.2004} marked by yellow points in \figref{fig_chiRPA}(a)].

In details, the lower branch of the ``hourglass'' can be interpreted as the resonance modes[shown in \figref{fig_chi}(a)] originating from local moments, influenced by itinerant spin modes[displayed in \figref{fig_chi}(b)] through the ``level repulsion'' of RPA correction, resulting in the transfer of spectral weight to lower energy around the $\boldsymbol{Q}_0$. It is essential to emphasize that the resonance mode at the exact $\boldsymbol{Q}_0$-point with characteristic energy $E_g$ remains protected without any spectral weight transfer. This protection results from the complete disappearance of the $a$-spinon dynamical spin susceptibility $\chi_a$ at this momentum due to the coherence factor effects discussed earlier. On the other hand, the spin fluctuation from fermionic itinerant $a$-spinons near $\boldsymbol{Q}_0$ is enhanced with the aid of that from local moments via the term $1-g^2 \chi_a(q) \chi_b(q)$ in RPA correction \eqnref{chiRPA}, leading to the upper branch in \figref{fig_chi}(b), which is relatively comparable to the lower branch primarily contributed by local moments. Additionally, the frequency slices of the calculated spin fluctuation spectrum for $\chi^{\mathrm{RPA}}$ around $\boldsymbol{Q}_0$ displayed in \figref{fig_chiRPA}(b)-(d) exhibit circular features deviating from $E_g$. This is distinct from the experimentally observed four weight peaks\cite{Greven.Chan.2016, Greven.Chan.2016noa, Keimer.Pailhès.2004, Keimer.Hinkov.2007, Tranquada.Xu.2009, Perring.Vignolle.2007, Dogan.Hayden.2004, Yamada.Tranquada.2004, Fujita.Sato.2020} marked by yellow points in \figref{fig_chiRPA})(d), suggesting that a higher-order correction might be needed to enhance them.

It is worth noting that all phenomenological parameters in our model include the resonance energy $E_g$, determined directly by the peak of weight in INS\cite{PhysRevLett.75.316, Keimer.He.20006wc, Bourges.Fauqu.2007, Keimer.Capogna.2007, Keimer.Fong.1999, Keimer.He.2002, Dai1999}, as well as $m_a$ and $\Delta_a$ for fermionic itinerant $a$-spinons, and the coupling strength $g$.  In this study, at $\delta=0.1$, we choose $2\Delta_a = 1.1 E_g$, $m_a= 1/J$, and $g=60\text{meV}$ to fit the experimental data, with $J=120 \text{meV}$ representing the bare spin exchange interaction. Also, the doping evolution of $m_a$ can be inferred from the relative change in the residual uniform spin susceptibility at low temperatures under strong magnetic fields\cite{Zheng.Kawasaki.2010j8}, the relationship with $m_a$ will be discussed in subsequent sections.  Furthermore, we show that the existence of the hourglass structure is insensitive to the specific choice of these parameters\cite{SM}, as long as the gap $2\Delta_a$ does not differ too much from the resonance energy $E_g$.

\begin{figure}[t]
\centering
\includegraphics[width=0.99\linewidth]{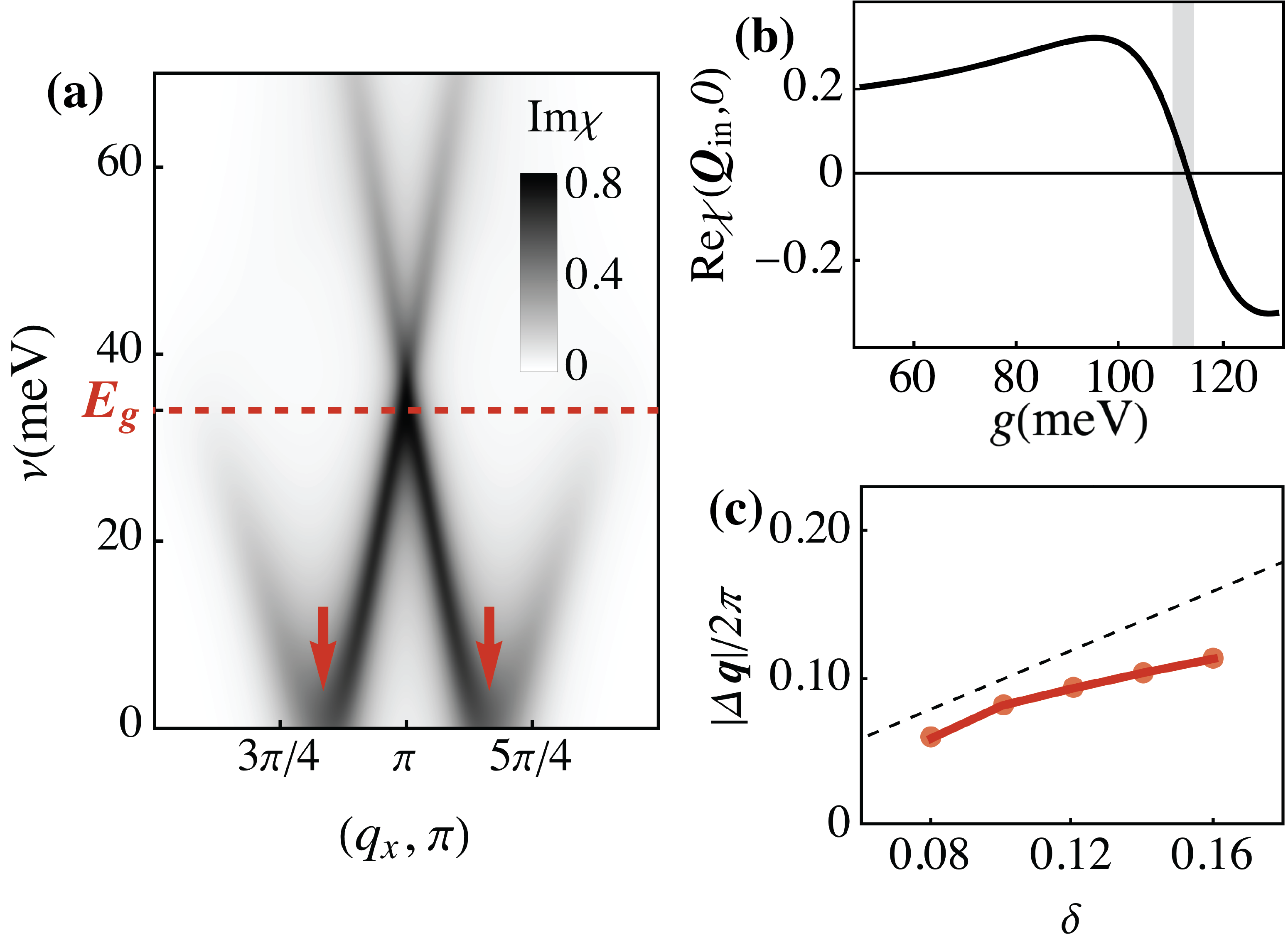}
\caption{(a) Calculated $\Im\chi^\text{RPA}(q)$ using \eqnref{chiRPA} at $\delta=0.1$ and $g=g_c$ displays gapless spin modes with incommensurate wave vector $\boldsymbol{Q}_\text{in}$ (red arrows). (b) Static spin susceptibility at $\boldsymbol{Q}_\text{in}$ determined by the real part of \eqnref{chiRPA}, showing sign change at $g=g_c$ (gray region). (c) Comparison of calculated doping evolution of $\Delta \boldsymbol{q}$ with experimental rule $2\pi\delta$ (dashed line).}
\label{fig_stripe}
\end{figure}

{\it Incommensurate magnetic instability.---} When the coupling strength $g$ approaches a critical value $g_c$,  sign changes in static susceptibility become possible, i.e., $\operatorname{Re} \chi^{\text{RPA}}(\omega=0, \boldsymbol{Q}_{\text{in}})<0$ as illustrated in \figref{fig_stripe}(b), at incommensurate momenta $\boldsymbol{Q}_{\text{in}}\equiv\boldsymbol{Q}_0 + \Delta\boldsymbol{q}$, alongside the gapless spin excitation shown in  \figref{fig_stripe}(a) stemming from the extension of the lower branch of the ``hourglass'' structure[with  $\boldsymbol{Q}_{\text{in}}$ marked by red arrows in \figref{fig_stripe}(a)]. This results in the emergence of incommensurate magnetic instability with wave vectors $\boldsymbol{Q}_{\text{in}}$, which may be associated with stripe order\cite{Uchida.Tranquada.1995, Matsuda.Fujita.2002, Gunnarsson.Zaanen.1989, Howald.Kivelson.2003, Chan.Zheng.2017, Devereaux.Huang.2017, Tranquada.Fradkin.2015doan, Yamada.Tranquada.2004} once circular gapless modes further break rotational symmetry and select a specific direction due to higher-order corrections.

Furthermore, the determination of the deviating incommensurate wave vector $\Delta\boldsymbol{q}$ for magnetic instability is related to the pocket size of itinerant $a$-spinon and the width of resonance modes, both of which increase with the rise in doping density $\delta$. As depicted in \figref{fig_stripe}(c), the doping evolution of $\Delta\boldsymbol{q}$ is consistent with experimental and theoretical conclusions\cite{Yamada.Tranquada.2004, Howald.Kivelson.2003}, i.e., $2\pi\delta$ as indicated by the dashed line.

{\it Unifrom susceptibility.---} The uniform static susceptibility in our study is contributed by both $a$-spinons and $b$-spinons, denoted as $\chi^{\text{loc}}=\chi_b^{\text{loc}}+\chi_a^{\text{loc}}$. Due to the existence of an energy gap for both $a$-spinons and $b$-spinons,  the uniform static susceptibility $\chi^{\text{loc}}$ appears to be significantly suppressed at temperatures close to zero. Nonetheless, in a specific situation where a strong magnetic field is applied,  it is possible to suppress $\Delta_a$ at the conventional vortex cores mediated by the emergent $U(1)$ gauge field between holons and $a$-spinons from the constraint \eqnref{Sab}\cite{SM}. Consequently, a finite DOS of $\mathcal{N}(0)=\frac{a^2}{2 \pi \hbar^2} m_a$ from the gapless Fermi pockets of $a$-spinon can be restored at these vortex cores, resulting in a finite residual $\chi_a^{\text{loc}}\propto \mathcal{N}(0)$ at low temperatures in cuprates, which is in agreement with the observed NMR results\cite{Zheng.Kawasaki.2010j8, Wen.Mei.2012}. Further details regarding the temperature evolution of $\chi^{\text{loc}}$ can be found in Ref. \onlinecite{SM}.

In addition, our previous work\cite{Zhang.Weng_2022,Weng.Ma.2014} suggests that the emergence of gapless $a$-spinon Fermi pockets when $\Delta_a$ is suppressed by strong magnetic fields can also account for the observed linear-$T$ heat capacity\cite{, Klein.Girod.2020, Wen.Mei.2012, Fang.Wen.2009} and the quantum oscillations\cite{Taillefer.Doiron-Leyraud.2007, Greven.Barii.2013} associated with pocket physics.

{\it Discussion.---}The hourglass-like spin excitation has been discussed as the consequence of a two-component RVB description of the doped Mott insulator at finite doping. Here two-component spinons characterize the local and itinerant spin moments emerging upon doping the single-band $t$-$J$ model, in contrast to the single-component spinon in the original RVB theory proposed by Anderson\cite{Anderson.Baskaran.1987, Anderson.Anderson.1987}. Note that the separation of itinerant spins (electrons) and local moments is a natural concept in multi-band systems such as the heavy fermion systems with Kondo coupling \cite{Levin.Auerbach.1986, Coleman.Coleman.1984, Fisk.Nakatsuji.2004} and iron-based superconductors with Hund's rule coupling \cite{Weng.Kou.2009, Ku.Yin.20101j9, Phillips.Lv.2010j0l, Teitel'baum.Gor'kov.2013, Weng.You.2011l2}, where the mutual interaction between the two degrees of freedom produces the correct low-lying spin excitations. In the present study, the emergence of two distinct spin components is due to the unique strong correlation effect within a single-band system that results in fractionalization. Specifically, the itinerant fermionic $a$-spinons carry the spin degrees of freedom associated with hopping holes, while the $b$-spinons describe the background local moments persisting from the half-filling. The interaction between these two components, as described in \eqnref{Sab}, arises from the no-double-occupancy constraint in the $t$-$J$ model.

In our study, the hourglass spectrum uniquely relies on the coherence factor effect \cite{Aksay.Fong.1995zl, Hu.Seo.2009, Eremin.Korshunov.2008, Scalapino.Maier.2008} of the $s$-wave pairing $\Delta^a$ of the itinerant spinons. It is worth pointing out that, within this framework (in the presence of holon condensation), the superconducting order parameters have a composition structure given by $\left\langle\hat{c}_{i \uparrow} \hat{c}_{j \downarrow}\right\rangle \propto \Delta_{ij}^a \langle e^{i \frac{1}{2}\left(\Phi_i^s+\Phi_j^s\right)}\rangle$, where the amplitude $\Delta^a$ is s-wave-like while the $d$-wave pairing symmetry as well as the phase coherence arise from the phase factor $e^{i \frac{1}{2}\left(\Phi_i^s+\Phi_j^s\right)}$ contributed by the $b$-spinons \cite{Weng.Ma.2014, Weng.Zhang.2020, Zhang.Weng_2022}. Such a hidden $s$-wave component with a BCS-like $d$-wave pairing order parameter leads to a novel pairing-symmetry dichotomy, which has been revealed and discussed in recent numerical\cite{Weng.Zhao.2022} and may have important experimental implications\cite{Xue.Zhong.2016, Feng.Ren.2016, Xue.Zhu.2021hwd}. Here the phase transition near $T_c$ is dictated by the free $b$-spinon excitations carrying the $\pi$-vortices \cite{SM,Weng.Ma.2014, Weng.Mei.20107w}.  %The $s$-wave pairing $\Delta_{ij}^a$ remains unaffected, maintaining the observed hourglass spectrum slightly above $T_c$. 
Finally, we shall show elsewhere how the spin excitations discussed in the present work may also naturally reduce to a commensurate AFM Goldstone mode in a dilute doping limit.

%\bibliography{revision_CP.bib}
%%%%%%
\begin{acknowledgments}
{\it Acknowledgments.---} We acknowledge stimulating discussions with Zhi-Jian Song, Zhen Bi, and Ji-Si Xu.
J.-X.Z., C.C., and Z.-Y.W. are supported by MOST of China (Grant No. 2017YFA0302902).
C.C. acknowledges the support from the Shuimu Tsinghua Scholar Program.
J.H.Z. is supported by a startup fund from the Pennsylvania State University (Zhen Bi), and thanks the hospitality of the Kavli Institute for Theoretical Physics, which is partially supported by the National Science Foundation under Grant No. NSF PHY-1748958.
\end{acknowledgments}

%%%%%% reference
\bibliography{draft_hg}

\clearpage
%%%%%%
\appendix
\widetext
\begin{center}
	\textbf{\large Supplementary Materials for: ``Hourglass-Like Spin Excitation in a Doped Mott Insulator''}
\end{center}

\vspace{1mm}

\renewcommand\thefigure{\thesection S\arabic{figure}}
\renewcommand\theequation{\thesection S\arabic{equation}}

\setcounter{figure}{0} 
\setcounter{equation}{0}

%%%%%%%%%%%%
%%%%%%
In the following supplementary materials, we provide more analytical results to support the conclusions presented in the main text. In Sec.~I., we present a detailed derivation of the dynamical spin susceptibility for itinerant fermionic $a$-spinons, $\chi_a(q)$, as given in \eqnref{chia}. In Sec.~II., we give the discrete energy levels for bosonic $b$-spinons, as well as a comprehensive derivation of the corresponding dynamical spin susceptibility $\chi_b(q)$ in \eqnref{chib}. In Sec.~III., we show that the four well-nested Fermi pockets of itinerant $a$-spinons, discussed in the main text, are consistent with the hopping fermions in the square lattice with uniform $\pi$-flux. In Sec.~IV., we reveal the existence of two types of vortex excitations in different temperature regions and provide the temperature evolution of spin susceptibility related to vortex states. In Sec.~V., we display the dynamical spin susceptibility $\chi^\text{RPA}(q)$ at the RPA level with various chosen parameters, illustrating that the ``hourglass'' feature is not sensitive to the specific parameters.

%%%%%

\section{I. Derivation of Dynamical Spin Susceptibility for Itinerant Fermionic $a$-Spinons in \eqnref{chia}}
Following the order of particle–hole and pocket degrees of freedom, we arrange the $a$-spinon operators as:
\begin{eqnarray}\label{psi}
    \psi_k&=&\left(\begin{array}{c}
        a_{k \uparrow} \\
        a_{-k \downarrow}^{\dagger}
        \end{array}\right) \otimes\left(\begin{array}{l}
        \Gamma \\
        X
    \end{array}\right)\\
	\Psi_k&=&\left(\begin{array}{c}
        a_{k \uparrow} \\
        a_{-k \downarrow}^{\dagger}
        \end{array}\right) \otimes\left(\begin{array}{l}
        M_x \\
        M_y
    \end{array}\right),
\end{eqnarray}
where $k=(i\omega_n, \boldsymbol{k})$ refers to the fermionic momentum-frequency vector. This work is primarily focused on the magnetic fluctuation around  $\boldsymbol{Q}_0=(\pi, \pi)$, thus only the particle-hole scattering between two pockets shifted by $\boldsymbol{Q}_0$ is relevant.  Specifically, scattering between $\Gamma$ and $X$ pockets, or $M_x$ and $M_y$ pockets, is considered. Consequently, the pocket indices consist of either $(\Gamma, X)$ or $(M_x, M_y)$ combinations. Using such representation, the Hamiltonian in \eqnref{Ha} can be written as
\begin{equation}
	H_a=\sum_{\boldsymbol{k}}\psi_{\boldsymbol{k}}^{\dagger} h_{\boldsymbol{k}} \psi_{\boldsymbol{k}} + \sum_{\boldsymbol{k}}\Psi_{\boldsymbol{k}}^{\dagger} h_{\boldsymbol{k}} \Psi_{\boldsymbol{k}}
\end{equation}
with
\begin{equation}
	h_k=\epsilon_{\boldsymbol{k}} \sigma_z \otimes \tau_0+\Delta_a \sigma_x \otimes \tau_0,
\end{equation}
where $\epsilon_K(\boldsymbol{k})=\boldsymbol{k}^2 / 2 m_a-\mu_a$ is the dispersion for $a$-spinons, and $\sigma$ and $\tau$ are Pauli matrices denoting the particle-hole and pocket degrees of freedom, respectively. Therefore, the Green's function for $a$-spinon $G_a(k) = -\left\langle\psi_k \psi_k^{\dagger}\right\rangle = -\left\langle\Psi_k \Psi_k^{\dagger}\right\rangle$ is given by:
\begin{eqnarray}
	G_a(k) \equiv\diag{Ga.png}{7pt}&=&\left(i\omega_n \sigma_0 \otimes \tau_0-h_{\boldsymbol{k}}\right)^{-1}\\
    &=&\frac{i\omega_n \sigma_0 \otimes \tau_0+\Delta_a \sigma_x \otimes \tau_0+\epsilon_{\boldsymbol{k}} \sigma_z \otimes \tau_0}{(i\omega_n)^2-E_{\boldsymbol{k}}^2},
\end{eqnarray}
where $E_{\boldsymbol{k}}=\sqrt{\epsilon_{\boldsymbol{k}}^2+\Delta_a^2}$ is the dispersion for $a$-spinon with BCS pairing. The dynamical spin susceptibility from itinerant $a$-spinons is defined as $\chi_a\left(r_i-r_j\right)=\left\langle S_a^z\left(r_i\right) S_{a}^z\left(r_j\right)\right\rangle$. $\chi_a$ can be expressed in the frequency-momentum space as follows:
\begin{equation}\label{chia0}
    \chi_a(q)=-2\times\frac{1}{4 N} \sum_k \operatorname{Tr} G_a(k+q) s_a G_a(k) s_a=\diag{chia.png}{24pt},
\end{equation}
where $s_a=\sigma_0 \otimes \tau_x$ and $s_a = \sigma_0 \otimes \tau_0$ denote the magnetic fluctuation near $(\pi, \pi)$ and $(0, 0)$, respectively. Note that the factor 2 in \eqnref{chia0} arises from the summation over $\psi$ and $\Psi$ components. Following the Matsubara summation, the expression for the dynamical spin susceptibility becomes \eqnref{chia}, which reads:
\begin{eqnarray}
    \chi_a(i v_{n},\boldsymbol{q})=-\frac{1}{2 N} \sum_{\boldsymbol{k}}\left(1-\frac{\Delta_a^2+\epsilon_{\boldsymbol{k}+\boldsymbol{q}} \epsilon_{\boldsymbol{k}}}{E_{\boldsymbol{k}+\boldsymbol{q}} E_{\boldsymbol{k}}}\right) \times \left(\frac{1}{i v_n-E_{\boldsymbol{k}+\boldsymbol{q}}-E_{\boldsymbol{k}}}-\frac{1}{i v_n+E_{\boldsymbol{k}+\boldsymbol{q}}+E_{\boldsymbol{k}}}\right),
\end{eqnarray}
where $\boldsymbol{q}$ represents the momentum deviation from $(0,0)$ and $(\pi,\pi)$. Furthermore, by replacing pocket indexes in \eqnref{psi} to $\left(\begin{array}{ll} \Gamma & M_y\end{array}\right)^T$ and $\left(\begin{array}{ll} \Gamma & M_x\end{array}\right)^T$, the dynamical spin susceptibility $\chi_a$ around $(\pi, 0)$ and $(0, \pi)$ can be determined, respectively. $\chi_a$ is found to be identical across all scenarios where $\boldsymbol{q}$ deviates from $(0, 0)$, $(0, \pi)$, $(\pi, 0)$, or $(\pi, \pi)$.

\section{II. Derivation of Dynamical Spin Susceptibility for Background Bosonic $b$-Spinons in \eqnref{chib}}
The $b$-spinons in the main text are in the RVB states on a square lattice under uniform magnetic flux. The corresponding Hamiltonian can be expressed as:
\begin{eqnarray}\label{Hb}
	H_b = -J_s \sum_{\langle ij \rangle, \sigma} b_{i\sigma}^\dagger b_{j-\sigma}^\dagger e^{i \sigma A_{ij}^h} + h.c. 
	+ \lambda_b \sum_{i,\sigma} ( b_{i\sigma}^\dagger b_{i\sigma} - N ),
\end{eqnarray}
Here the assumed gauge field $A_{ij}^h$ comes from the mutual Chern-Simons interaction between holons and background $b$-spinons. Therefore, with the holons condensed, the RVB-pairing $b$-spinons experience a uniform static gauge field with a $\delta\pi$ flux per plaquette.

\begin{figure}[tb]
	\centering
	\includegraphics[scale=0.8]{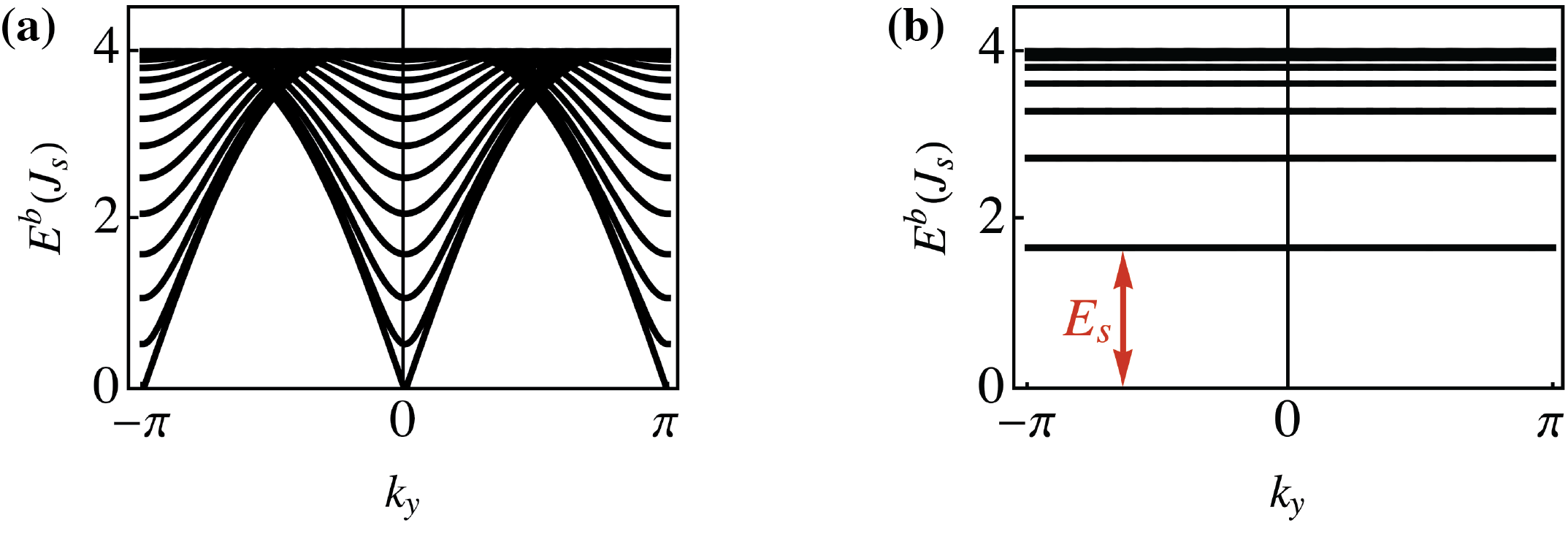}
	\caption{The dispersion of $b$-spinons $E_m$ from \eqnref{Emb} with all quantum numbers plotted along $k_y$ for (a) zero flux and (b) uniform $\delta\pi$ flux in minimum square plateaus. Red arrow indicates lowest excitation level $E_s$. Parameters: $\delta=1/8$ and $\lambda_b=4J_s$.}
	\label{fig_Eb}
\end{figure}

Now, the pairing component can be redefined as:
\begin{align}
    \sum_{i,j} b_{i,\uparrow}^\dagger M_{i,j} b_{j,\downarrow}^\dagger + b_{i,\downarrow} M_{i,j} b_{j,\uparrow},
\end{align}
where $M$ is a hermitian matrix defined as:
\begin{equation}\label{Mmat}
        M_{i,j} =
    \begin{cases}
    -J_s e^{i A_{ij}^h} & j \in \text{NN}(i) \\
    0 & \text{others}
    \end{cases}
\end{equation}
Then, with the standard diagonalization procedure as in Hofstadter system, we obtain:
\begin{equation}
	H_{b}=\sum_{m, \sigma} E_{m}^b \gamma_{m \sigma}^{\dagger} \gamma_{m \sigma}
\end{equation}
with the $b$-spinons spectrum:
\begin{equation}\label{Emb}
	E_{m}^b=\sqrt{\lambda_b^{2}-(\xi_{m}^b)^{2}}
\end{equation}
via introducing the following Bogoliubov transformation:
\begin{equation}\label{bogo}
	b_{i \sigma}=\sum_{m} \omega_{m \sigma}(\boldsymbol{r}_i)\left(u_{m} \gamma_{m \sigma}-v_{m} \gamma_{m-\sigma}^{\dagger}\right),
\end{equation}
where the coherent factors are given by
\begin{eqnarray}\label{Bogofactor}
u_{m}&=&\sqrt{\frac{1}{2}\left(1+\frac{\lambda}{E_{m}^b}\right)}
\notag\\
	v_{m}&=&\operatorname{sgn}\left(\xi_{m}^b\right) \sqrt{\frac{1}{2}\left(-1+\frac{\lambda}{E_{m}^b}\right)}.
\end{eqnarray}
Here, $\xi_{m}^b$ as well as $w_{m}(\boldsymbol{r}_i)\equiv w_{m \sigma}(\boldsymbol{r}_i)=w_{m-\sigma}^{*}(\boldsymbol{r}_i)$ in \eqnref{bogo} are the eigenfunctions and eigenvalues of the following equation:
\begin{equation}\label{Hbdiag}
	\xi_{m}^b \omega_{m }(\boldsymbol{r}_i)=-\frac{J \Delta^{s}}{2} \sum_{j=\text{NN}(i)} e^{i \sigma A_{i j}^{h} }\omega_{m }(\boldsymbol{r}_j).
\end{equation}

We select the Landau gauge along the $x$-axis, as expressed in  $A_{i,i+\hat{e}_y}^h=-\delta \pi i_x$. The resulting $b$-spinon dispersion $E_m^b$ in \eqnref{Emb} with the unit of $J_s$  is depicted in \figref{fig_Eb}(b), which manifests the dispersionless, ``Landau-level-like'' discrete energy levels\cite{Zhang.Weng_2022, Weng.Chen.2005} with a gap $E_s$[labeled by the red arrow]. For comparison, \figref{fig_Eb}(a) displays the continuous spectra for conventional Schwinger bosons under zero flux conditions, highlighting the low-lying propagating modes.  For the sake of clear representation, we depict all quantum numbers excluding $k_y$ simultaneously in the figures.

%where $J_s=J_{\text{eff}}\Delta^s/2$ and $J_{\text{eff}}=J(1-\delta)^2-2\gamma\delta^2$. In this scheme, the parameters $\lambda_b$ and $\Delta^s$ are determined by self-consistent calculation \eqnref{self2}. 
\begin{figure}[tb]
	\centering
	\includegraphics[scale=0.7]{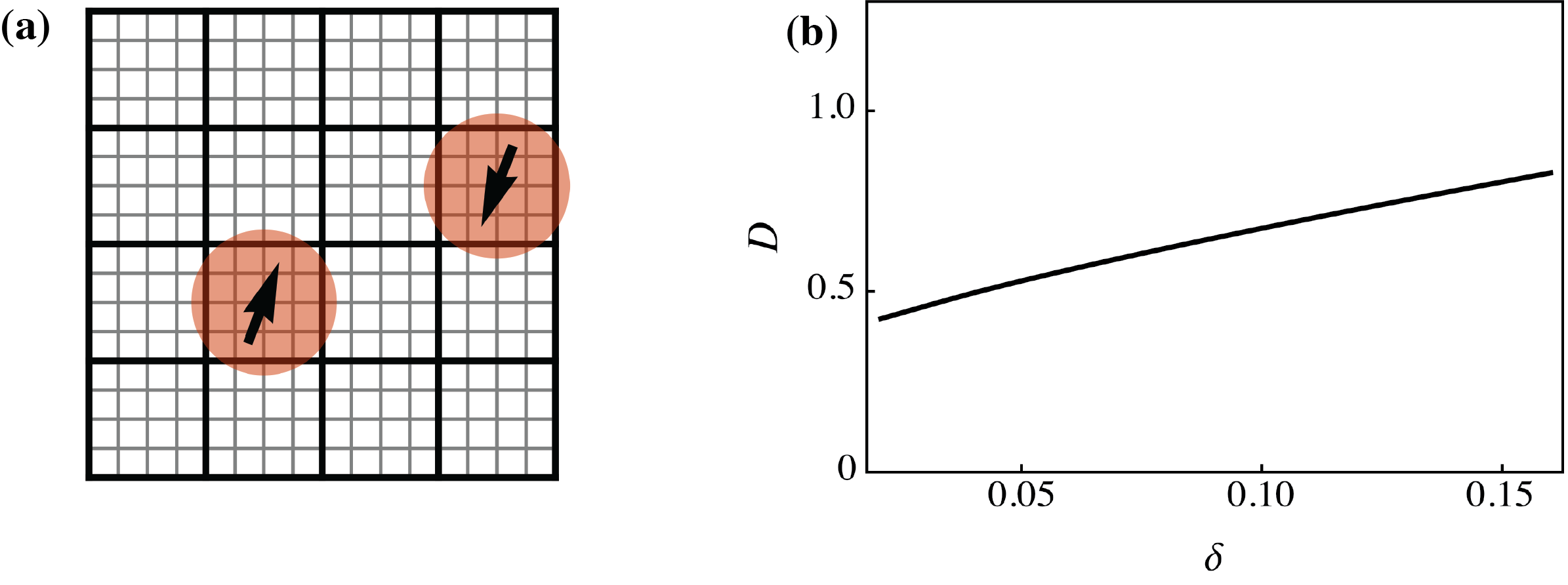}
	\caption{(a) Depicts $b$-spinon wave packets, labeled by red disks, with either $w_{m^*}(\mathbf{r})$ or $w_{m}(\mathbf{r})$ magnetic Wannier wave functions in a square lattice (gray grids). These are positioned within individual magnetic unit cells (black grids).  (b) The doping evolution of $D$ in \eqnref{DD} as calculated in prior mean-field self-consistent studies\cite{Weng.Ma.2014, Zhang.Weng_2022}.}
	\label{fig_D}
\end{figure}

Subsequently, using the relation $S_{i}^{b, z}=\frac{1}{2} \sum_{\sigma} \sigma b_{i \sigma}^{\dagger}b_{i \sigma}$, the Matsubara spin-spin correlation function can be expressed as:
\begin{eqnarray}\label{chib1}
&\;&\chi_b(\tau, \boldsymbol{r}_i- \boldsymbol{r}_j) =\left\langle\hat{T} S_{j}^{b, z}(\tau) S_{i}^{b, z}(0)\right\rangle_{0}
\\
	&=&\frac{1}{4} \sum_{\sigma \sigma^{\prime}} \sigma \sigma^{\prime}\left\langle\hat{T} b_{j \sigma}^{\dagger}(\tau) b_{j \sigma}(\tau) b_{i \sigma^{\prime}}^{\dagger}(0) b_{i \sigma^{\prime}}(0)\right\rangle_{0} \\
	&=&\frac{1}{4} \sum_{\sigma \sigma^{\prime}} \sigma \sigma^{\prime}\left[\left\langle\hat{T} b_{j \sigma}^{\dagger}(\tau) b_{i \sigma^{\prime}}^{\dagger}(0)\right\rangle_0\left\langle\hat{T} b_{j \sigma}(\tau) b_{i \sigma^{\prime}}(0)\right\rangle_0 + \left\langle\hat{T} b_{j \sigma}^{\dagger}(\tau) b_{i \sigma^{\prime}}(0)\right\rangle_0\left\langle\hat{T} b_{j \sigma}(\tau) b_{i \sigma^{\prime}}^{\dagger}(0)\right\rangle_0\right]
\end{eqnarray}
where $\langle\rangle_0$ denotes the expectation value under the mean-field state, and the Wick's theorem is applied in the last line. Then, by using the Bogoliubov transformation \eqnref{bogo}, together with the Green's function
\begin{eqnarray}
	G_{\gamma}\left(m, i \omega_{n} ; \sigma\right) &\equiv& -\left\langle\gamma_{m \sigma}(i\omega_n) \gamma_{m \sigma}^{\dagger}(i\omega_n)\right\rangle_{0}\notag\\
	&=&\frac{1}{i \omega_{n}-E_{m}^b}
\end{eqnarray}
After performing the summation over $\sigma$ and replacing $w_{m,\sigma}$ with $w_m$, the Matsubara spin correlation function in \eqnref{chib1} at $T=0$ can be further simplified as:
\begin{eqnarray}\label{chib2}
	\chi_b(i \nu_n, \boldsymbol{r}_i-\boldsymbol{r}_j) &=& -\frac{1}{4} \sum_{m,n} w_{m}^*(\boldsymbol{r}_i)w_m(\boldsymbol{r}_j) w_{n}^*(\boldsymbol{r}_j) w_{n}(\boldsymbol{r}_i)
	(u_m^2 v_n^2 + v_m^2 u_n^2 - 2 u_m v_m u_n v_n) \notag \\
	&\;&\times \left( \frac{1}{i \nu_n - E_m - E_n} - \frac{1}{i \nu_n + E_m + E_n} \right)
\end{eqnarray}

From the second line of \eqnref{chib2}, the dominant contribution to $\chi_b$ evidently originates from the lowest Landau level(LLL), wherein $E_m = E_n = E_s$, leading to $u_m = u_n$ according to \eqnref{Bogofactor}. Thus,
the only non-vanishing contributions are from the cases where $v_n = -v_m$, i.e., $\xi_n^b = -\xi_m^b$. Moreover, according to previous works\cite{Weng.Mei.20107w, Efros.Rashba.1997}, in the LLL, there exists $N_m$ eigenvectors of $M$ matrix in \eqnref{Mmat} ($N_m$ is the number of magnetic unit cells), with $w_{m}(\boldsymbol{r})$ peaking at the center of a magnetic unit cell located at $\mathbf{R}_m$. We term these localized $w_m$'s as \emph{local} modes (LM). 
Moreover, for each local mode $w_m(\boldsymbol{r})$, a corresponding eigenvector $w_{m^*}(\boldsymbol{r}) = (-1)^{\boldsymbol{r}} w_{m}(\boldsymbol{r})$ exists, and $\xi_{m^*}^b = -\xi_m^b$, therefore $u_{m^*} = u_m$ and $v_{m^*} = -v_m$. We term these $w_{m^*}$'s as \emph{$\pi$-shifted} modes.

The Bogoliubov quasiparticles corresponding to both local and $\pi$-shifted modes possess a common energy $E_s$ and a common $u_m$, but $v_m$ differs in sign between these two classes of modes. In essence, under this approximation, the low-lying spin spectrum $\chi_b$ will be dominated by the localized $b$-spinon excitations, which are non-propagating modes with an intrinsic size on the order of a ``cyclotron length'', $a_c$. These spinon wave packets with magnetic Wannier wave functions $w_{m^*}(\mathbf{r})$ or $w_{m}(\mathbf{r})$ are situated in separate magnetic unit cells and are highly degenerate, as illustrated in \figref{fig_D}(a).

In the summation of \eqnref{chib2}, $m$ and $n$ will be either local or $\pi$-shifted modes, thus we find:
\begin{eqnarray}
	\chi_b(i \nu_n, \boldsymbol{r}_i-\boldsymbol{r}_j) &=&  \frac{1}{4}  (-1)^{\boldsymbol{r}_i-\boldsymbol{r}_j} 2 \left| \sum_{m \in \text{LM}} w_m^*(\boldsymbol{r}_i) w_m(\boldsymbol{r}_j) \right|^2 \left(1-\frac{4 \lambda_b^2}{E_g^2} \right)\left( \frac{1}{i \nu_n - E_g} - \frac{1}{i \nu_n + E_g} \right) \nonumber \\
	& =&  \frac{1}{4} (-1)^{\boldsymbol{r}_i - \boldsymbol{r}_j} e^{-\frac{1}{2a_c^2}(\boldsymbol{r}_i - \boldsymbol{r}_j)^2} \frac{1}{2\pi^2 a_c^4} \left(1-\frac{4 \lambda_b^2}{E_g^2} \right)\left( \frac{1}{i \nu_n - E_g} - \frac{1}{i \nu_n + E_g} \right).
	\end{eqnarray}
where $E_g=2E_s$ is the resonance energy discussed in the main text. Here, we employ the fact that:
\begin{equation}
		\left|\sum_{m \in \text{LM}} w_{m}^*(\boldsymbol{r}) w_m(\boldsymbol{r}') \right| = \frac{1}{2\pi a_c^2} e^{-(\boldsymbol{r} - \boldsymbol{r}')^2/4a_c^2},
	\end{equation}
where $a_c = 1/\sqrt{\pi \delta}$ is the cyclotron length, and we assume lattice constants to be unit, i.e., $a=1$ for simplicity. By executing a Fourier transformation into the momentum space, we can obtain the expression in \eqnref{chib}:
\begin{eqnarray}\label{chib3}
	\chi_b(i\nu_n,  \boldsymbol{Q}) &= &\diag{chib.png}{4pt} = \frac{1}{N} \sum_{\boldsymbol{r}} \chi_b(i\nu_n,  \boldsymbol{r}) e^{-i \boldsymbol{Q} \cdot \boldsymbol{r}}\notag \\
	&= & \frac{1}{4}\frac{1}{\pi  a_c^2} \left(1-\frac{4\lambda_b^2}{E_g^2}\right) e^{-\frac{a_c^2}{2}(\boldsymbol{Q} - \boldsymbol{Q}_0)^2}
	\left( \frac{1}{i \nu_n - E_g} - \frac{1}{i \nu_n + E_g} \right) \\
    &= & a_c^2 D e^{-\frac{a_c^2}{2}\left(\boldsymbol{Q}-\boldsymbol{Q}_0\right)^2}  \times \left(\frac{1}{i\nu_n-E_g}-\frac{1}{i\nu_n+E_g}\right), \notag
	\end{eqnarray}
where $\boldsymbol{Q}_0 = (\pi,\pi) $ is the AFM wave vector, and $D$ is defined as:
\begin{equation}\label{DD}
    D\equiv\frac{1}{4} \frac{1}{\pi a_c^4}\left(1-\frac{4 \lambda_b^2}{E_g^2}\right).
\end{equation}
The doping dependence of the weight of $\chi_b$ is mainly contributed from $a_c^2$ in the last line of \eqnref{chib3}, rather than from the value of $D$.  \figref{fig_D}(b) shows the doping evolution of $D$ based on the mean-field self-consistent calculation from the prior work\cite{Weng.Ma.2014, Zhang.Weng_2022}, demonstrating the insensitivity of $D$ with respect to the doping density $\delta$.

\section{III. BCS States of Fermions in a Square Lattice with Uniform $\pi$-Flux}
\begin{figure}[tb]
	\centering
	\includegraphics[width=\linewidth]{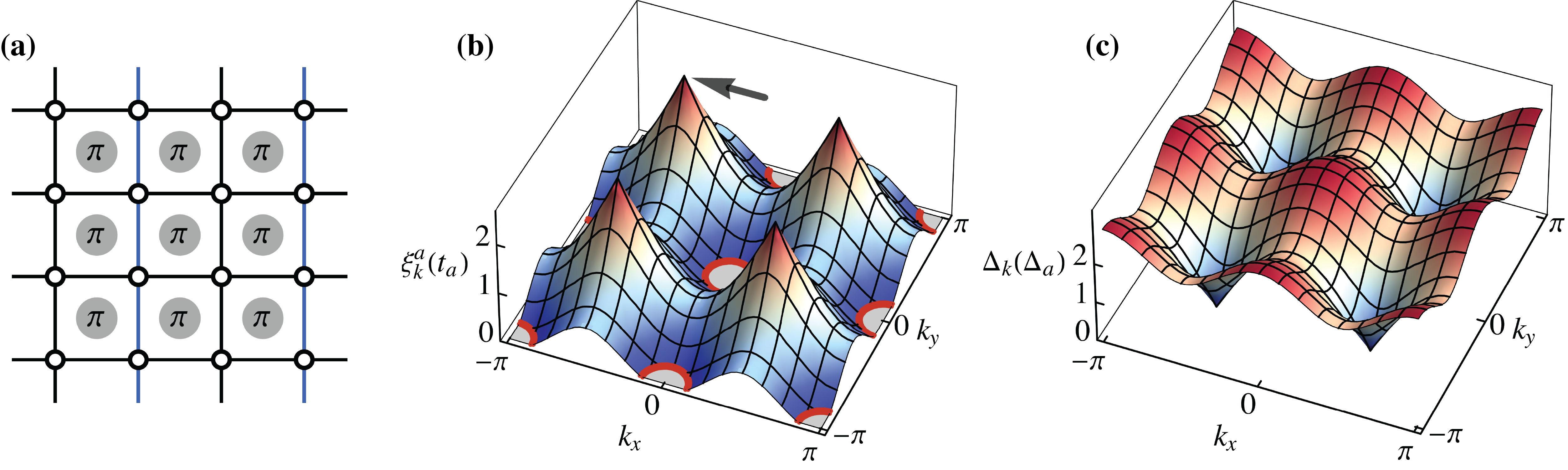}
	\caption{(a) Square lattice illustration with uniform $\pi$-flux. One possible gauge choice for $\phi_{i j}^0$ is shown, with black bonds for nearest-neighbor links ($e^{i\phi_{i j}^0}=+1$) and blue bonds for links with $e^{i\phi_{i j}^0}=-1$. (b) The dispersion $\xi_{\boldsymbol{k},-}$ (from \eqnref{piEa}) of free fermions in a square lattice with uniform $\pi$-flux. The red arrow indicates the well-known Dirac point for half-filling ($n=1$). Red circles denote Fermi pockets at particle density $n=\delta$ ($\delta=0.1$). (c) Pairing order $\Delta_k$ from \eqnref{piDelta}.}
	\label{fig_Ea}
\end{figure}

Assume that fermions form nearest-neighbor (NN) pairing on a square lattice with uniform $\pi$ flux, as depicted in \figref{fig_Ea}(a). The Hamiltonian for this setup is provided in 
\begin{equation}\label{Hasup}
	H_\pi=-t_a \sum_{\langle i j\rangle, \sigma} a_{i \sigma}^{\dagger} a_{j \sigma} e^{-i \phi_{i j}^0}- \Delta_a\sum_{\langle i j\rangle, \sigma}  \sigma a_{i \sigma} a_{j \bar{\sigma}} e^{i \phi_{i j}^0}+\text { h.c. } +\lambda_a\left(\sum_{i, \sigma} a_{i \sigma}^{\dagger} a_{i \sigma}-\delta N\right),
\end{equation}
where $\phi_{ij}^0$ is the $\pi$-flux gauge field, while $\Delta_a$ and $\lambda_a$ denote the NN pairing amplitude and the chemical potential, respectively. The latter constrains the number of fermions to equal that of doping holes. Selecting the Landau gauge displayed in \figref{fig_Ea}(a) yields the dispersion of \eqnref{Hasup} as presented in
\begin{equation}
E_{k,\pm}=\sqrt{\left(\xi_{k,\pm}\right)^2+\Delta_k^2},
\end{equation}
where $\xi_{k, \pm}$ and $\Delta_k$ is the dispersion for free fermions and the $s$-wave BCS type pairing order parameter, as specified in
\begin{eqnarray}
	\xi_{k, \pm}&=& \pm 2 t_a \sqrt{\cos ^2 k_x+\cos ^2 k_y}+\mu_a \label{piEa} \\
	\Delta_k&=&2 \Delta_a \sqrt{\cos ^2 k_x+\cos ^2 k_y}.\label{piDelta}
\end{eqnarray}
The lower branch dispersion $\xi_{k,-}$ from \eqnref{piEa} is portrayed in \figref{fig_Ea}(a), exhibiting well-nested Fermi pockets denoted by red circles. Here, we can understand the origin of this gapless ``Fermi pockets'' as follows: according to the \eqnref{Hasup}, in the absence of pairings, free fermions are in the $\pi$-flux lattices, of which the half-filled case corresponds to the well-known $\pi$-flux state in fermionic spin liquids, with the Fermi surface shrinking to the Dirac point marked by the red arrow in \figref{fig_Ea}(a). However, the number of fermions corresponds to the doping density $\delta$, not half-filling, which results in the Dirac point transforming into a gapless Fermi pocket, as illustrated by the red circles in \figref{fig_Ea}(a).

Furthermore, the calculated BCS type pairing order parameter $\Delta^k$ is shown in \figref{fig_Ea}(b), demonstrating a strongly momentum-dependent $s$-wave without sign flip. As our focus lies on the physics near the Fermi surface of $\xi_{k,-}$—namely, around $(0,0)$, $(\pi,0)$, $(0,\pi)$, and $(\pi,\pi)$—the anisotropy of the pairing amplitude is not crucial.

Finally, the next-nearest neighbor (NNN) hopping term solely opens the gap of Dirac points, as indicated by the red arrows in \figref{fig_Ea}(a), implying that such further neighbor term would not affect the Fermi pockets, which are our primary concern at low energy. As a result, coupled with the features of well-nested pockets and the $s$-wave pairing presented in \eqnref{Hasup}, hopping fermions on the square lattice with uniform $\pi$-flux emerge as a potential model. This model could account for the low-lying physical behaviors of $a$-spinons discussed in the main text. Moreover, the mean-field phase string theory in earlier work\cite{Weng.Ma.2014, Zhang.Weng_2022} can provide the effective Hamiltonian \eqnref{Hasup}.

\section{IV. Vortex Types and Temperature Evolution of Uniform Spin Susceptibility}
In phase string theory, we identify two distinct types of "vortices" generated by the magnetic fields. Specifically, under holon condensation, the experimentally observed superconducting order parameters are given by:
\begin{equation}\label{cc}
  \left\langle\hat{c}_{i \uparrow} \hat{c}_{j \downarrow}\right\rangle \propto \Delta_{ij}^a e^{i \frac{1}{2}\left(\Phi_i^s+\Phi_j^s\right)}, 
\end{equation}
with the $d$-wave pairing symmetry arising from the phase $e^{i \frac{1}{2}\left(\Phi_i^s+\Phi_j^s\right)}$, which is contributed by $b$-spinons\cite{Weng.Ma.2014, Weng.Zhang.2020, Zhang.Weng_2022}.

The magnetic field induces a novel magnetic $\pi$-vortex core which entraps a free $b$-spinon, and suppresses RVB pairing $\Delta_s$, while $\Delta_a$ remains unaffected [illustrated in Figure 1(b)]. This gives rise to a phase transition near $T_c$, manifesting Kosterlitz-Thouless-like behavior\cite{Weng.Ma.2014, Weng.Mei.20107w}. This behavior disrupts only the phase $e^{i \frac{1}{2}\left(\Phi_i^s+\Phi_j^s\right)}$ in \eqnref{cc} due to the novel magnetic $\pi$-vortices.

%The first is the novel magnetic $\pi$-vortex core, trapping a free $b$-spinon and suppressing RVB pairing $\Delta_s$ while maintaining $\Delta_a$ unchanged [as illustrated in \figref{fig_uniform}(b)].

\begin{figure}[tb]
	\centering
	\includegraphics[scale=0.8]{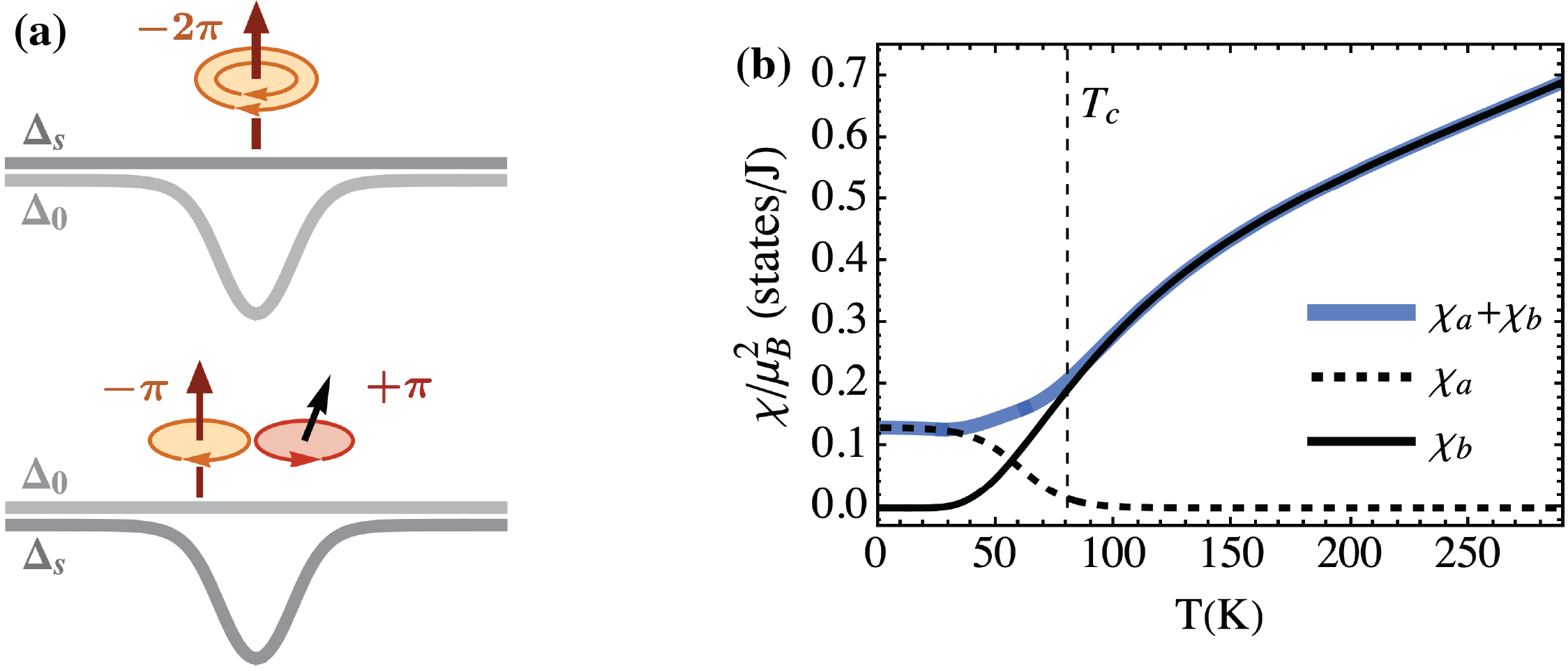}
	\caption{Illustration for  two types of vortices under a magnetic field in (a) and (b). (a)Conventional $2\pi$ vortex trapping $-2\pi$ external magnetic flux[labeled by red arrows], with $\Delta_a$ suppressed at the vortex core but $\Delta_s$ is preserved.  (b)Novel $\pi$ vortex trapping $-\pi$ external magnetic flux [denoted by the red arrow], along with a free $b$-spinon [indicated by the black arrow]. In this case, $\Delta_s$ is suppressed at the vortex core, while $\Delta_a$ is preserved.  (c)The temperature evolution of static uniform spin susceptibility, with black solid line denoting $\chi^{\mathrm{loc}}_b$ in \eqnref{chibuni}, with black dashed line denoting $\chi^{\mathrm{loc}}_a$ in \eqnref{chia2}, with blue line denoting $\chi^{\mathrm{loc}}$ in \eqnref{totchi}.Notably, the variations in $\chi^{\mathrm{loc}}$ under the magnetic field are discernible when comparing $\chi^{\mathrm{loc}}_b$ (black solid line) and $\chi^{\mathrm{loc}}$ (blue line).}
	 \label{fig_uniform}
\end{figure}

On the other hand, there also exists the conventional magnetic vortex with a quantization of $2\pi$. This causes the phase of $\Delta_a$ in  \eqnref{cc} to twist, resulting in the unpairing of $a$-spinons at the vortex cores mediated by the emergent $U(1)$ gauge field, which comes from the constraint \eqnref{Sab}. In contrast, $b$-spinons remain gapped[illustrated in \figref{fig_uniform}(b)].

The two vortex types appear within distinct temperature domains. At temperatures much lower than $E_g/k_B$, novel magnetic $\pi$-vortices may be energetically unfavorable due to the minimum $b$-spinon gap $E_s=E_g/2$ required to break an RVB pair. This is in contrast to a conventional $2\pi$-vortex where $\Delta_a=0$. However, near $T_c$, $\pi$-vortices carrying $b$-spinons are more readily formed under external magnetic fields, preceding the disruption of superconducting phase coherence by thermally excited spinon-vortices.

Furthermore, our study investigates the contribution of both $a$-spinons and $b$-spinons to the uniform static susceptibility $\chi^{\mathrm{loc}}$, as expressed by
\begin{equation}\label{totchi}
 \chi^{\mathrm{loc}}=\chi_b^{\mathrm{loc}}+\chi_a^{\mathrm{loc}}.
\end{equation}
To derive the uniform static susceptibility $\chi_b^{\mathrm{uni}}$ for $b$-spinons, we introduce the external magnetic field in \eqnref{Hb}, represented as $-2 \mu_B \sum_i S_i^z H$. This inclusion leads to the Zeeman splitting effect in the $b$-spinon dispersion given by:
\begin{equation}
    E_{m,\sigma}^b=E_m^b-\sigma \mu_B H,
\end{equation}
where $E_m^b$ is defined in \eqnref{Emb}. Consequently, the total magnetic moment induced by the magnetic field from $b$-spinons can be expressed as:
\begin{equation}
    M_b=\mu_B \sum_m\left[n_B\left(E^b_{m,\uparrow}\right)-n_B\left(E^b_{m,\downarrow}\right)\right]
\end{equation}
where $n_B(\omega)=1 /\left(e^{\beta \omega}-1\right)$ denotes the bosonic distribution function. Therefore, the $\chi_b^{\mathrm{loc}}$ at local site is defined by $\chi_b^{\mathrm{loc}} = \frac{M_b}{N B} \mid_{H \rightarrow 0}$, resulting in
\begin{equation}\label{chibuni}
    \chi_b^{\text{loc}}=\frac{2 \beta\mu_B^2}{N}\sum_m n_B(E_m)[n_B(E_m)+1],
\end{equation}
The temperature evolution of $\chi_b^{\text{uni}}$ as described in \eqnref{chibuni} is depicted by the black solid line in \figref{fig_uniform}(c). It can be observed that $\chi_b^{\text{uni}}$ decreases as the temperature decreases due to the strengthening antiferromagnetic correlations, which oppose the uniform polarization of the spin. Moreover, the existence of an energy gap in $b$-spinons leads to the opening of a gap at low temperatures, approximately below $T_c$. The values of the parameters used in our calculations are determined by the mean-field self-consistent equations presented in Ref. \onlinecite{Weng.Ma.2014, Zhang.Weng_2022}.

Furthermore, the uniform static susceptibility $\chi_a^{\text{loc}}$ for $a$-spinons can be derived by setting $\boldsymbol{q}\rightarrow 0$ and $\mu \rightarrow 0$ in \eqnref{chia}, resulting in
\begin{eqnarray}\label{chia2}
    \chi_a^{\text{uni}}=\frac{2}{N} \sum_{\boldsymbol{k}} n_F\left(E_{\boldsymbol{k}}\right)\left[1-n_F\left(E_{\boldsymbol{k}}\right)\right]
\end{eqnarray}
where $n_F(\omega)=1 /\left(e^{\beta \omega}+1\right)$ denotes the fermionic distribution function. At low temperature, $\chi_a^{\text{loc}}$ in \eqnref{chia2} can be further simplified as a temperature-independent Pauli susceptibility directly related to the density of states (DOS) $\mathcal{N}(0)$ at the Fermi surface. However, itinerant fermionic $a$-spinons possess a BCS-type gap $\Delta_a$, leading to the disappearance of $\mathcal{N}(0)$ and uniform static susceptibility $\chi_a^{\text{loc}}$.  However, itinerant fermionic $a$-spinons possess a BCS-type gap $\Delta_a$, resulting in the disappearance of $\mathcal{N}(0)$ and the uniform static susceptibility $\chi_a^{\text{loc}}$. Nevertheless, the application of a strong magnetic field can suppress $\Delta_a$ at conventional $2\pi$ vortex cores, leading to the restoration of a finite DOS with $\mathcal{N}(0)=\frac{a^2}{2 \pi \hbar^2} m_a$ contributed by the gapless Fermi pockets of $a$-spinons. This restoration induces a finite residual $\chi_a^{\text{loc}}$ given by:
\begin{equation}
    \chi_a^{\text{loc}} = 2 \mathcal{N}(0) = 2 \frac{a^2}{2 \pi \hbar^2} m_a F(T)
\end{equation}
where an additional coefficient $F(T)$ is introduced to account for the temperature effect of conventional $2\pi$-vortex, which only exists below temperature $T_c$. The specific expression of $F(T)$ is irrelevant for the structure of $\chi_a^{\text{loc}}$, and for simplicity of representation, we assume $F(T) = [\exp[\left(T-0.75 T_c\right) / 0.1T_c]+1]^{-1}/4$. Therefore, the black dashed line in \figref{fig_uniform}(c) represents $\chi_a^{\text{loc}}$.

As the result, the total uniform static susceptibility $\chi^{\mathrm{loc}}$ in \eqnref{totchi} under a strong magnetic field, is depicted by the blue line in \figref{fig_uniform}(c). Comparing it with the case without magnetic fields, in which $\chi_a^{\text{loc}}$ vanishes completely [shown by the black solid line in \figref{fig_uniform}(c)], we observe the emergence of a finite residual $\chi^{\mathrm{loc}}$ under a strong magnetic field when $T<T_c$. This finding is consistent with the NMR measurements.

\section{V. Comparison of $\Im \chi^\text{RPA}$ in \eqnref{chiRPA} for Various Parameters}

\begin{figure}[tb]
	\centering
	\includegraphics[width=\linewidth]{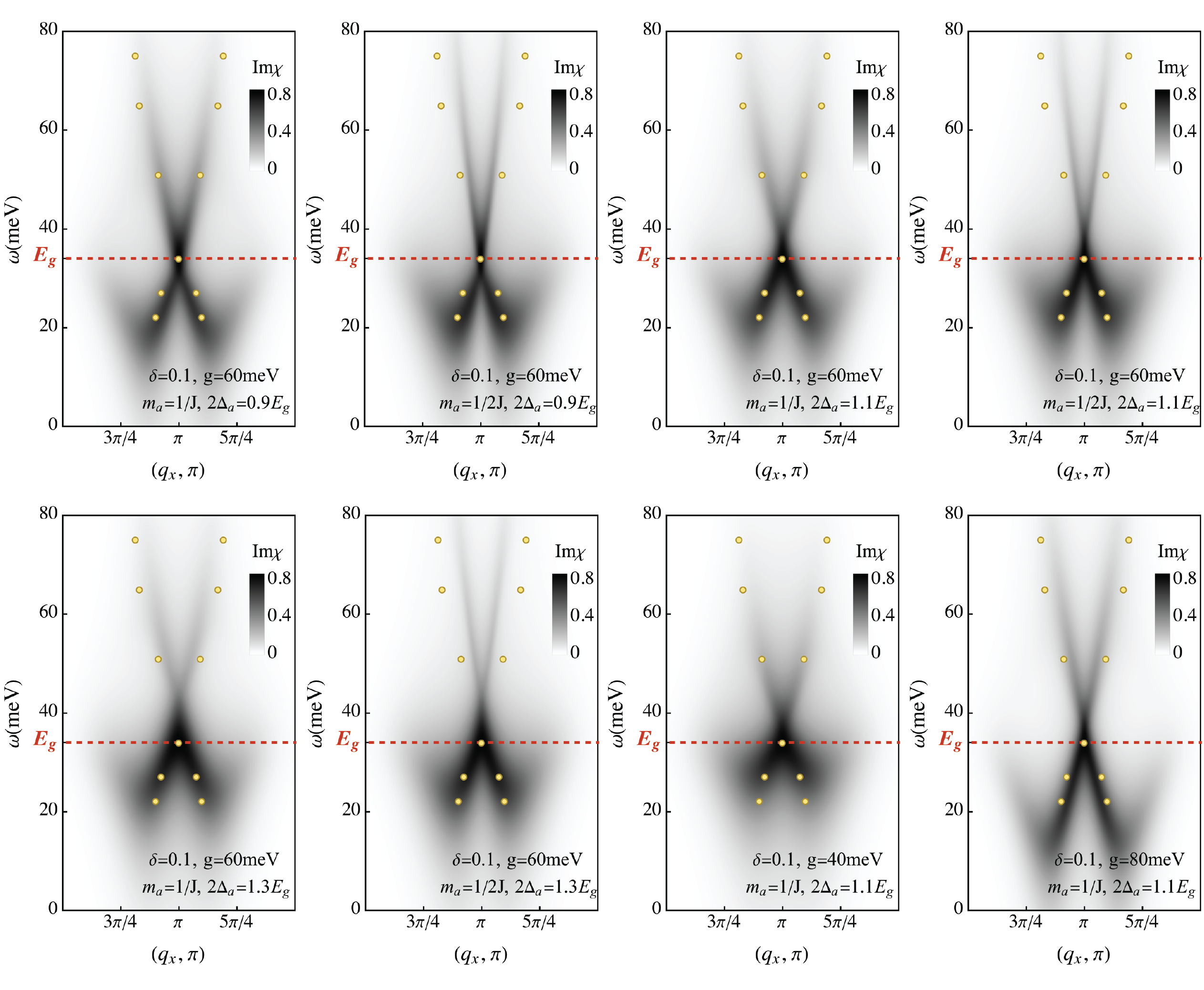}
	\caption{ Imaginary part of dynamic spin susceptibility at RPA level, $\Im\chi^\text{RPA}(q)$, determined by \eqnref{chiRPA} around AFM wave vector $\boldsymbol{Q}_0$ at $\delta=0.1$ with varying parameters.}
	\label{fig_fitparameter}
\end{figure}

In the main text, we select specific values for the fitting parameters at $\delta=0.1$, namely $2\Delta_a = 1.1 E_g$, $m_a= 1/J$, and $g=60\text{meV}$, to match the experimental data, where $J=120 \text{meV}$ represents the bare spin exchange interaction. Furthermore, we present the results of $\Im\chi^\text{RPA}(q)$ determined by \eqnref{chiRPA} for different parameter choices at $\delta=0.1$ in \figref{fig_fitparameter}. These results clearly demonstrate that the presence of the hourglass structure is not significantly affected by the specific values of these parameters, as long as the gap $2\Delta_a$ is not too different from the resonance energy $E_g=2E_s$. This condition is reasonable because the BCS-type pairing $\Delta_a$ for $a$-spinons originates from the RVB pairing $\Delta_s$ for $b$-spinons, following the relation $\left|\Delta_a\right|^2 \simeq \delta^2\left|\Delta_s\right|^2$.

\end{document}